%% file: main.tex
\documentclass[a4paper,UKenglish]{lipics-v2018}

\usepackage{microtype}

\usepackage{makeidx}  
\usepackage{amsmath,amssymb}
\usepackage{soul}
\usepackage{times}
\usepackage{graphicx}
\usepackage[ruled,vlined]{algorithm2e}
\usepackage{gastex} 
\usepackage{listings}

\newcommand{\cost}{\mbox{\rm R}}

\newcommand{\Exp}{\mathbb{E}}

\newcommand{\Z}{\ensuremath{{\rm \mathbb Z}}}

\newcommand{\ExpRew}{\mathsf{ExpRew}}

\newcommand{\act}{A}
\newcommand{\mov}{\Gamma}
\newcommand{\trans}{\delta}
\newcommand{\stra}{\sigma}
\newcommand{\bigstra}{\Sigma}
\newcommand\distr{{\mathcal D}}
\newcommand\pat{\pi}
\newcommand\pats{\Pi}
\newcommand{\game}{G}
\newcommand{\supp}{\mathrm{Supp}}

\newcommand{\cala}{{\mathcal A}}
\newcommand{\Avg}{\mathsf{Avg}}
\newcommand{\LimInfAvg}{\mathsf{LimInfAvg}}
\newcommand{\LimSupAvg}{\mathsf{LimSupAvg}}
\def\set#1{\{ #1 \}}

\newcommand{\ov}{\overline}

\newenvironment{compactitem}{\begin{itemize}}{\end{itemize}}

\newenvironment{compactenum}{\begin{enumerate}}{\end{enumerate}}

\title{Ergodic Mean-Payoff Games for the Analysis of Attacks in Crypto-Currencies}

\author{Krishnendu Chatterjee}{IST Austria}{krishnendu.chatterjee@ist.ac.at}{}{}

\author{Amir Kafshdar Goharshady}{IST Austria}{amir.goharshady@ist.ac.at}{}{}

\author{Rasmus Ibsen-Jensen}{IST Austria}{ribsen@ist.ac.at}{}{}

\author{Yaron Velner}{Hebrew University of Jerusalem}{yaron.welner@mail.huji.ac.il}{}{}

\authorrunning{K. Chatterjee, A.K. Goharshady, R. Ibsen-Jensen and Y. Velner}

\Copyright{Krishnendu Chatterjee, Amir K. Goharshady, Rasmus Ibsen-Jensen and Yaron Velner}

\subjclass{Software and its Engineering $\rightarrow$ Software Verification and Validation $\rightarrow$ Formal Software Verification}


\category{}

\relatedversion{}

\supplement{}

\funding{}

\EventShortTitle{CONCUR 2018}
\EventAcronym{CONCUR}
\EventYear{2018}
\nolinenumbers 
\hideLIPIcs  

\begin{document}

\maketitle

\input{abstract}

\input{introduction}

\input{background}
\input{definitions}

\input{modeling}

\input{short_formal_modeling}

\input{experimental_results}
\input{related}

\input{conclusion}



\bibliography{concur}
\clearpage
\input{appendix}

\input{formal-modeling}
\input{app_exper}

\end{document}

%% file: abstract.tex
\begin{abstract}
Crypto-currencies are digital assets designed to work as a medium of exchange,
e.g., Bitcoin, but they are susceptible to attacks (dishonest behavior of participants).
A framework for the analysis of attacks  in crypto-currencies requires 
(a)~modeling of game-theoretic aspects to analyze incentives for deviation 
from honest behavior;
(b)~concurrent interactions between participants; and 
(c)~analysis of long-term monetary gains.
Traditional game-theoretic approaches for the analysis of security protocols 
consider either qualitative temporal properties such as safety and termination, 
or the very special class of one-shot (stateless) games. 
However, to analyze general attacks on protocols for crypto-currencies, 
both stateful analysis and quantitative objectives are necessary.
In this work our main contributions are as follows:
(a)~we show how a class of concurrent mean-payoff games, namely ergodic 
games, can model various attacks that arise naturally in 
crypto-currencies; 
(b)~we present the first practical implementation of algorithms for ergodic 
games that scales to model realistic problems for crypto-currencies; 
and (c)~we present experimental results showing that our framework can 
handle games with thousands of states and millions of transitions.
\keywords{Cryptocurrency, Bitcoin, Ergodic Games, Quantitative Analysis}
\end{abstract}

%% file: introduction.tex
\section{Introduction}

\noindent{\textbf{Economic effects of security violations.}}
Traditionally, automated security analysis of protocols using game-theoretic
frameworks focused on qualitative properties, such as safety or liveness~\cite{KremerR03,CR14,abadi2000reconciling},
to ensure absolute security.
In many cases absolute security is too expensive, and 
security violations are inevitable.
In such scenarios rather than security, the economic implications 
of violations should be accounted for.
In general, economic consequences of security violations are hard to 
measure. However, there is a new application area of 
crypto-currencies, in which the economic impact of an attack can be measured in terms of the number of coins that are lost. These currencies have considerable market value, in the order of hundreds of billions of dollars \cite{coinmarketcap},
thus developing a framework to formally analyze the security violations and their 
economic consequences for crypto-currencies is an interesting problem.

\smallskip\noindent{\textbf{Crypto-currencies.}}
There are many active crypto-currencies today, some with considerable market values. Currently, the main crypto-currency is Bitcoin with a value of over 150 billion dollars at the time of writing \cite{coinmarketcap}. Virtually all of these currencies are free from outside governance and authority and are not controlled by any central bank. Instead, they work based on the decentralized \emph{blockchain} protocol. This protocol, which was first developed for monetary transactions in Bitcoin \cite{nakamoto2008bitcoin}, sets down the rules for creating new units of currency and valid transactions. However, it only defines the outcomes of actions taken by involved parties and cannot dictate the actions themselves. So, the whole ecosystem operates in a game-theoretic manner. The lack of an authority also leads to irreversibility of transactions, so if an amount of currency is transferred unintentionally or due to a bug, it cannot be reclaimed. This, together with the huge market values, makes it imperative to develop formal methods for quantifying the economic consequences before deploying the protocols.

\smallskip\noindent \textbf{Dishonest interaction.}
The fact that protocols define only the outcomes of actions (in terms of loss or earning of currency), and do not force the actions themselves, means that in some scenarios they might give one of the parties unfair or unintended advantage over others and an incentive to act dishonestly, i.e.~to take an unintended action. 
Such behavior is called an attack.
We succinctly describe some attacks.
\begin{compactitem}
\item The most fundamental attack in every crypto-currency is 
 \emph{double-spending}, where one party could in some circumstances 
use the same coin twice in two different purchases.
While this vulnerability is inherent in every blockchain protocol, people 
still use crypto-currencies as the probability (and the economic consequences)
of such an attack can be bounded over time.

\item Another line of attacks follow from dishonest behavior of 
the \emph{blockchain miners} who are responsible for the underlying 
security of the blockchain protocol and are rewarded for their operations.
It was shown that undesirable behavior, such as block withholding~\cite{eyal2015miner} 
or selfish mining~\cite{EyalS14}, could increase the dishonest miner's reward, 
at the expense of other (honest) miners.
We explain the block withholding attack in more detail in Section~\ref{subsub:bwh}.
\end{compactitem}

\smallskip\noindent\textbf{Research Questions.}
Analyzing attacks on crypto-currencies requires a formal framework to handle:
(a)~game-theoretic aspects and incentives 
for dishonest behavior; 
(b)~simultaneous interaction of the participants; and 
(c)~quantitative properties corresponding to long-term monetary gains and losses. These properties cannot be obtained from standard temporal or qualitative properties which have been the focus of previous game-theoretic frameworks~\cite{KremerR03,CR14}.
On the other hand, game-theoretic incentives are also analyzed in the security 
community~(e.g., see~\cite{bonneau2015sok}), but their methods are normally considering the very 
special case of one-shot (stateless) or short-term games.
One-shot games cannot model the different states of the ecosystem or the history of actions taken by participants.

\smallskip\noindent\textbf{Concurrent mean-payoff games.}
These games were introduced in the seminal
work of Shapley~\cite{Sha53}, and later extended by Gillette~\cite{Gil57}.
A concurrent mean-payoff game is played by two players over a finite state 
space, where at each state both players simultaneously choose actions.
The transition to the next state is determined by their joint actions, and 
each transition is assigned a reward. 
The goal of one player is to maximize the long-run average of the rewards, and
the other player tries to minimize it.
These games provide a very natural and general framework to study stateful 
games with simultaneous interactions and quantitative objectives. 
They lead to a very elegant and mathematically rich 
framework, and the theoretical complexity of such games has been studied 
for six decades~\cite{Sha53,Gil57,BF68,HK66,MN81,CMH08,HKLMT11}. 
However, the analysis of concurrent mean-payoff games is computationally 
intractable and 
no practical (such as strategy-iteration) algorithms exist to solve 
these games. 
Existing algorithmic approaches either require the theory of reals and 
quantifier elimination~\cite{CMH08} or have doubly-exponential time complexity 
in the number of states~\cite{HKLMT11}, and cannot handle beyond toy examples 
of ten transitions.

\smallskip\noindent\textbf{Our contributions.}
Our main contributions are as follows:
\begin{compactenum}
\item \textbf{Modeling.}
We propose to model long-term (infinite-horizon) economic aspects of security violations as 
concurrent mean-payoff games, between the attacker and the defender.
The guaranteed payoff in the game corresponds to the maximal loss of the
defender.
In particular, for blockchain protocols, where the utility of every transition
is naturally measurable, we show how to model various
interesting scenarios as a sub-class of concurrent mean-payoff games,
namely, {\em concurrent ergodic games}.
In these games all states are visited infinitely often with probability~1.

\item \textbf{Practical implementation.} 
Second, while for concurrent ergodic games a theoretical algorithm (strategy-iteration 
algorithm) exists that does not use theory of reals and quantifier elimination, 
no previous implementation exists. 
Moreover, the implementation of the theoretical algorithm 
poses practical challenges: (a)~the algorithm guarantees convergence
only in the limit; and (b)~the algorithm requires high numerical precision and 
the straightforward implementation of the algorithm does not converge 
in practice.
We present (i)~a simple stopping criterion for approximation,
and (ii)~resolve the numerical precision problem; and to our knowledge present
the first practical implementation of a solver for concurrent ergodic games.

\item \textbf{Experimental results.}
Finally, we present experimental results and show that the solver for ergodic games scales to 
thousands of states and nearly a million transitions to model realistic analysis problems 
from crypto-currencies.
Note that in comparison, approaches for general concurrent mean-payoff games 
cannot handle even ten transitions (see the Remark in Section~\ref{rem:inefficiency}).
Thus we present orders of magnitude of improvement.
\end{compactenum}

%% file: background.tex
\section{Crypto-Currencies}

\noindent \textbf{Monetary system.}
A crypto-currency is a \emph{monetary system} that allows secure transactions 
of currency units and dictates how new units are formed.
Each transaction has a unique id and the following components:
(i)~a set of inputs;  and 
(ii)~a set of outputs and (iii)~locking scripts. 
Each input has a pointer to an output of a previous transaction, and 
each output has an assigned monetary value. A locking script on an output defines a condition for using the funds stored in that output, e.g. the need for a digital signature. An input can use funds of the output it points to only if it can satisfy this condition. 

\smallskip\noindent\textbf{Validity.}
A transaction is {\em valid} if these conditions hold:
(a) the total value brought by the inputs is greater than or equal to the total value of the outputs; (b) the inputs have not been spent before; (c) the inputs satisfy locking scripts.

Note that the list of transactions is the only state of the system and higher level concepts like account balance and users are computed directly from it.
A transaction-based system is not secure if transactions are sent directly between users to transfer units.
While validity conditions are enough to make sure that only valid recipients could redirect units they once truly held, there is nothing in the transactions themselves to limit the user from spending the same output twice (in two different transactions).
For this purpose a public ledger of all valid transactions, called a  \emph{blockchain},  is maintained. 

\begin{figure}[!tb]
\begin{center}



\includegraphics[scale=0.2]{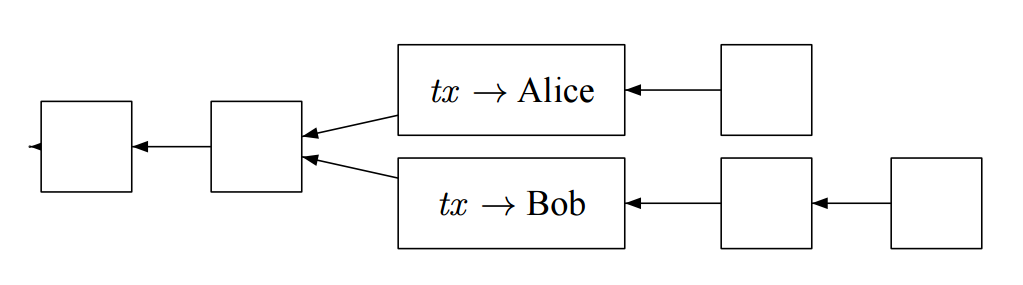}

\caption{The longest chain dictates that the transaction $\mathit{tx}$ belongs to Bob.}\label{fig:fork}
\end{center}

\end{figure}

\smallskip\noindent \textbf{Blockchain.}
A ledger is a distributed database that maintains a growing \emph{ordered} list of \emph{valid} transactions.
Its main novelty is that it enforces consensus among untrusted and possibly adversarial parties \cite{nakamoto2008bitcoin}.
In Bitcoin (and most other major crypto-currencies) the public ledger is implemented as a series of \emph{blocks} of transactions, each containing a reference to its previous block, and is hence called a blockchain.
A consensus on the chain is obtained by a decentralized pseudonymous protocol.
 Any party tries to collect new transactions, form a block and add it to the chain (this process is called block mining).
However, in order to do so, they must solve a challenging computational puzzle (which depends on the last block of the chain).
The process of choosing the next block is as follows:
\begin{compactenum}
\item The first announced valid block that solves the puzzle is added to the chain.
\item If two valid blocks are found approximately at the same time (depending on network latency), then there is a temporary fork in the chain. 
\end{compactenum}
Every party is free to choose either fork, and try to extend it.
Hence, the underlying structure of the blockchain is a tree.
At any given time, the longest path in the tree, aka the \emph{longest chain}, is the consensus blockchain 
(see Figure~\ref{fig:fork}).
Due to the random nature of the computational puzzle one branch will eventually become strictly longer than the other, and all parties will adopt it.

\smallskip\noindent \textbf{Mining process.}
The puzzle asks for a block consisting of valid transactions, hash of the previous block and an arbitrary integer $\mathit{nonce}$, whose hash is less than a target value.
The random nature of the hash function dictates a simple strategy for mining: try random nonces until a solution is found.
So the chance of a miner to find the next block is proportional to their computational power.

\smallskip\noindent\textbf{Incentives for mining.}
There are two incentives for miners: (i) Every transaction can donate to the miner who finds a new block that contains it, (ii) Each block
creates a certain number of new coins which are then given to the miner.

\smallskip\noindent \textbf{Pool mining.}
To lower the variance of their revenue, miners often collaborate in \emph{pools}~\cite{rosenfeld2011analysis,bonneau2015sok}.
The pools have a manager who collects the rewards from valid blocks found by the members and allocates funds to them in proportion to the amount of work they did.
Members prove their work by sending \emph{partial solution} blocks, which are blocks with valid transactions but lower difficulty level, i.e., the hash of the block is not smaller than the network threshold, but it is lower than some threshold that was defined by the manager.
As a result, pool members obtain lower variance in rewards, but have a small drop in expected revenue to cover the manager's fee.
Members will get the same reward for a partial and full solution, but the member cannot claim the full block reward for themselves. More precisely, a block also dictates where the block reward goes to. Hence, even if a member broadcasts the new block, the reward will still go to the manager.

\smallskip\noindent \textbf{Overview.}
A crypto-currency is a network with nodes. Some of the nodes are also miners.
A node has a local copy of the blockchain and local \emph{transaction pool}, which holds valid pending transactions that are still not in the blockchain.
When a user performs a transaction his associated nodes broadcast the transaction to the network.
When a node receives a new transaction it checks whether it is valid wrt its blockchain and transaction pool.
When a node receives a new block, it verifies that it is valid wrt consensus chain. If it is valid it adds it to the chain and updates his transaction pool accordingly.
Whenever a new valid transaction or block is received, the node broadcasts it to all of its neighbors.

\smallskip\noindent \textbf{Proof of stake mining.}
An emerging criticism over the huge amount of energy that is wasted in the mining process led to development of \emph{proof of stake protocols}.
In proof of stake mining the miner is elected with probability that is proportional to their \emph{stake} in the network (i.e., number of coin units he holds), rather than their computation power.
Current proof of stake protocols assume a synchronous setting~\cite{nxt,casper,consmos} where a miner is chosen in every time slot $t_0$.
However, they differ in the way they reach consensus.
We study a simplified version of~\cite{consmos}.
\begin{compactenum}
\item At time $t_0$ a miner is randomly elected. She broadcasts the next block.
\item Until time $t_0 + t$ other miners who receive the block, verify it and if it were valid, sign it and broadcast the signature.
\item The block is added to the chain only if a majority of the network sign it.
\end{compactenum}
To encourage honest behavior, the elected miner and signers get rewards when the suggested block is accepted. 

%% file: definitions.tex
\newcommand{\rand}{r}
\newcommand{\Rec}{\mathsf{Rec}}
\newcommand{\BestResponse}{\mathsf{BestResponse}}
\newcommand{\ImpSw}{\mathsf{ImpSw}}
\newcommand{\OneStep}{\mathsf{OneSt}}

\section{Concurrent and Ergodic Games}\label{sec:def}

We first present the basic definitions and results related to concurrent
games.

\smallskip\noindent{\bf Probability distributions.}
For a finite set~$A$, a {\em probability distribution\/} on $A$ is a
function $\trans\!:A\to[0,1]$ such that $\sum_{a \in A} \trans(a) = 1$.
We denote the set of probability distributions on $A$ by $\distr(A)$. 
Given a distribution $\trans \in \distr(A)$, we denote by $\supp(\trans) = 
\{x\in A \mid \trans(x) > 0\}$ the {\em support\/} of the distribution. 

\smallskip\noindent{\bf Concurrent game structures.} 
A {\em concurrent stochastic game structure\/} 
$\game =  (S, \act,\mov_1, \mov_2, \trans)$ has the 
following components:
\begin{compactitem}
\item A finite state space $S$ and a finite set $\act$ of actions (or moves).

\item Two move assignments $\mov_1, \mov_2 \!: S\to 2^{\act}
	\setminus \emptyset$.  For $i \in \{1,2\}$, assignment
	$\mov_i$ associates with each state $s \in S$ the non-empty
	set $\mov_i(s) \subseteq \act$ of moves available to Player~$i$
	at state $s$.  

\item A probabilistic transition function
	$\trans\!:S\times\act\times\act \to \distr(S)$, which
	associates with every state $s \in S$ and moves $a_1 \in
	\mov_1(s)$ and $a_2 \in \mov_2(s)$, a probability
	distribution $\trans(s,a_1,a_2) \in \distr(S)$ for the
	successor state.
\end{compactitem}

We denote by $n$ the number of states (i.e., $n=|S|$), and by 
$m$ the maximal number of actions available for a player at a state 
(i.e., $m=\max_{s\in S} \max\set{|\mov_1(s)|,|\mov_2(s)|}$). 
The size of the transition relation of a game structure is defined as
$|\delta|=\sum_{s\in S}\sum_{a_1 \in \mov_1(s)} \sum_{a_2 \in \mov_2(s)} | \supp(\trans(s,a_1,a_2))| \leq n^2\cdot m^2$.

\smallskip\noindent{\bf Plays.}
At every state $s\in S$, Player~1 chooses a move $a_1\in\mov_1(s)$,
and simultaneously and independently
Player~2 chooses a move $a_2\in\mov_2(s)$.  
The game then proceeds to the successor state $t$ with probability
$\trans(s,a_1,a_2)(t)$, for all $t \in S$. 
A {\em path\/} or a {\em play\/} of $\game$ is an infinite sequence
$\pat =\big( (s_0,a^0_1, a^0_2), (s_1, a^1_1, a^1_2), (s_2,a_1^2,a_2^2)\ldots\big)$ of states and action pairs such that for all 
$k\ge 0$ we have (i)~$a^k_i \in \mov_i(s_k)$; and 
(ii)~$s_{k+1} \in \supp(\trans(s_k,a^k_1,a^k_2))$.
We denote by $\pats$ the set of all paths.

\begin{example}
	Consider a repetitive game of rock-paper-scissors, consisting of an infinite number of laps, in which
	each lap is made of a number of rounds as illustrated in Figure \ref{fig:rps}. 
	When a lap begins, the two players play rock-paper-scissors repetitively until one of them wins $3$ rounds more than her opponent, in which case she wins the current lap of the game and a new lap begins.
	In each round, the winner is determined by the usual rules of rock-paper-scissors,~i.e. rock beats scissors, scissors beat paper and paper beats rock. In case of a tie, each player wins the round with probability $\frac{1}{2}$.
	
	Here we have $S = \{ -2, -1, 0, 1, 2\}$ and $\Gamma_1 = \Gamma_2 \equiv \{\rm R, \rm P, \rm S\}$. The game starts at state $0$ and state $s$ corresponds to the situation where Player~$1$ has won $s$ rounds more than Player~$2$ in the ongoing lap. Edges in the figure correspond to possible transitions in the game. Each edge is labeled with three values $a_1, a_2, p$ to denote that the game will transition from the state at the beginning of the edge to the state at its end with probability $p$ if the two players decide on actions $a_1$ and $a_2$, respectively. For example, there is an edge from state $2$ to state $0$ labeled $\rm R, \rm S, 1$, which corresponds to
	$\delta(2, \rm R, \rm 
	S)(0) = 1$. 
	In the figure, we use $X, X$ in place of $a_1, a_2$ to denote that they are equal. Hence every \emph{play} in this game corresponds to an infinite walk on the graph in Figure \ref{fig:rps}.

	\begin{figure}[!hb]
		\begin{center}
		\includegraphics[scale=0.025]{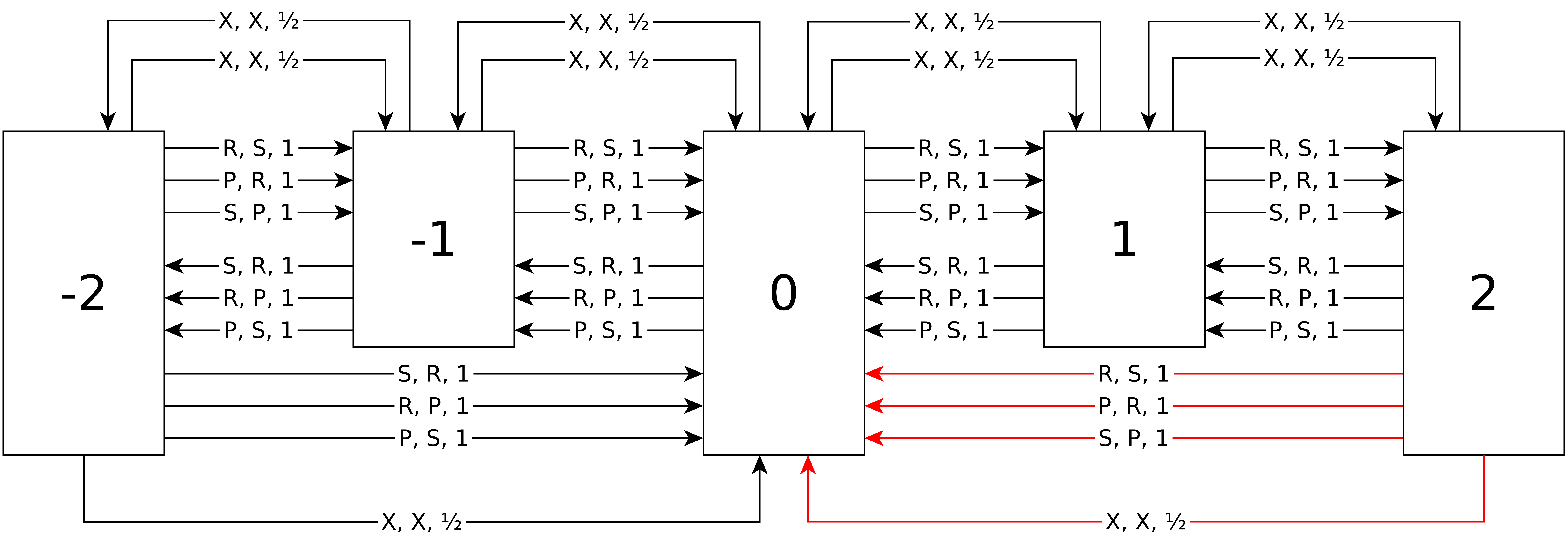}

		\caption{A repetitive rock-paper-scissors game}
		\label{fig:rps}
	\end{center}
\vspace{-8mm}
	\end{figure}	
\end{example}

\noindent{\bf Strategies.}
A {\em strategy\/} is a recipe to extend 
prefixes of a play.
Formally, a strategy for Player~$i$ is a mapping 
$\stra_i\!:(S\times \act \times \act)^* \times S \to\distr(\act)$ that 
associates with every finite sequence $x \in (S\times \act \times \act)^*$  
of state and action pairs, representing 
the past history of the game, and the current state $s$ in $S$, a probability distribution $\stra_i(x \cdot s)$ 
used to select the next move. 
The strategy $\stra_i$ can only prescribe moves that are available to 
Player~$i$; that is, for all sequences $x\in (S \times \act \times \act)^*$ 
and states $s\in S$, we require $\supp(\stra_i(x\cdot s)) \subseteq \mov_i(s)$. 
We denote by $\bigstra_i$ the set of all strategies for Player $i$. 
Once the starting state $s$ and the strategies $\stra_1$ and $\stra_2$
for the two players have been chosen, 
then the probabilities of measurable events are uniquely defined~\cite{VardiP85}. 
For an event $\cala\subseteq\pats$, we denote by 
$\Pr_s^{\stra_1,\stra_2}(\cala)$ the probability that a path belongs to 
$\cala$ when the game starts from $s$ and the players use the strategies 
$\stra_1$ and~$\stra_2$; and $\Exp_{s}^{\stra_1,\stra_2}[\cdot]$ 
is the expectation measure.
We call a pair of strategies $(\stra_1,\stra_2)\in \bigstra_1 \times\bigstra_2$ 
a \emph{strategy profile}.

\smallskip\noindent{\bf Stationary (memoryless) and positional strategies.} 
In general, strategies use randomization, and can use finite or even infinite 
memory to remember the history. 
Simpler strategies, that either do not use memory, or randomization, or both, 
are significant, as they are simple to implement and interpret.
A strategy $\sigma_i$ is \emph{stationary} (or memoryless) if it is 
independent of the history but only depends on the current state, i.e., 
for all $x,x'\in(S\times A\times A)^*$ and all $s\in S$, we have 
$\sigma_i(x\cdot s)=\sigma_i(x'\cdot s)$, and thus can be expressed as a 
function $\stra_i: S \to \distr(\act)$. 
A strategy is \emph{pure} if it does not use randomization, 
i.e., for any history there is always some unique action $a$ that is played 
with probability~1.
A pure stationary strategy $\stra_i$ is called {\em positional}, and 
represented as a function $\stra_i: S \to \act$.


\smallskip\noindent{\bf Mean-payoff objectives.}
We consider maximizing \emph{limit-average} (or mean-payoff) objectives for 
Player~1, and the objective of Player~2 is the opposite (i.e., the games are 
zero-sum). 
We consider concurrent games with a reward function $\cost: S\times\act \times 
\act \to \mathbb{R}$ that assigns a reward value $\cost(s,a_1,a_2)$ 
for all $s\in S$, $a_1 \in \mov_1(s)$, and $a_2 \in \mov_2(s)$.
For a path $\pat= \big((s_0, a^0_1, a^0_2), (s_1, a^1_1,a^1_2), \ldots\big)$, 
the average for $T$ steps is $\Avg_T(\pat)= \frac{1}{T} \cdot \sum_{i=0}^{T-1} \cost(s_i,a^i_1,a^i_2)$,
and the limit-inferior average (resp. limit-superior average) is defined as 
follows:
$\LimInfAvg(\pat)= \lim\inf_{T \to \infty} \Avg_T(\pat)$ 
(resp. $\LimSupAvg(\pat)= \lim\sup_{T \to \infty} \Avg_T(\pat)$).
For brevity we denote concurrent games with mean-payoff objectives as CMPGs 
(concurrent mean-payoff games).

\begin{example}
	Consider the game in Figure \ref{fig:rps}. In this game, Player $1$ wins a lap whenever a red edge is crossed. Therefore, in order to capture the number of laps won by Player $1$, rewards can be assigned as:
	$R(2, R, S) = R(2, P, R) = R(2, S, P) = 1; ~
	R(2, X, X)= \frac{1}{2}
	$ and $0$ in all other cases.
	
\end{example}

\noindent{\bf Values and $\epsilon$-optimal strategies.}
Given a CMPG $G$ and a reward function $\cost$, the \emph{lower value} 
$\underline{v}_s$ (resp. the \emph{upper value} $\ov{v}_s$) at a state $s$ 
is defined as follows:
\vspace{-2mm}
\[
\underline{v}_s =  \sup_{\stra_1 \in \bigstra_1 } \inf_{\stra_2 \in \bigstra_2} 
\Exp_s^{\stra_1,\stra_2}[\LimInfAvg]; 
\qquad 
\ov{v}_s  =  \inf_{\stra_2 \in \bigstra_2} \sup_{\stra_1 \in \bigstra_1} 
\Exp_s^{\stra_1,\stra_2}[\LimSupAvg].
\]
The {\em determinacy}  result of~\cite{MN81} shows that the upper 
and lower values coincide and give the \emph{value} of the game denoted as 
$v_s$.
For $\epsilon\geq 0$, a strategy $\stra_1$ for Player~1 is 
\emph{$\epsilon$-optimal} if we have $v_s - \epsilon \leq 
\inf_{\stra_2\in \bigstra_2} \Exp_s^{\stra_1,\stra_2}[\LimInfAvg]$.

\smallskip\noindent{\bf Ergodic Games.}  A CMPG $G$ is \emph{ergodic} if for all states 
$s,t \in S$, for all strategy profiles $(\stra_1,\stra_2)$, if we start at 
$s$, then $t$ is visited infinitely often with probability~1 in the 
random walk $\pat_s^{\stra_1,\stra_2}$.
The game in Figure \ref{fig:rps} is not ergodic. If Player~$1$ keeps playing rock and Player~$2$ scissors, then the states $-1$ and $-2$ are visited at most once each. We now present a more realistic version of the same game that is also ergodic.

\begin{example}
	Consider two players playing the repetitive game of rock-paper-scissors over a network, e.g.~the Internet. The game is loaded on a central server that asks the players for their moves and provides them with rewards and information about changes in the state of the game. Given that the network is not perfect, there is always a small probability that one of the players is unable to announce his move in time to the server. In such cases, the player will lose the current round. Assume that this scenario happens with probability $\epsilon > 0$. 	Then all probabilities in Figure \ref{fig:rps} have to be multiplied by $(1 - \epsilon)$ and new transitions, which are not under players' control and are a result of uncertainty in the network connection, should be added to the game. These new transitions are illustrated in Figure \ref{fig:netrps}. Here a star can be replaced by any permissible action of the players. It is easy to check that this variant of the game is ergodic, given that starting from any state, there is a positive probability of visiting any other state within $3$ steps using the new transitions only.

			\begin{figure}[!hb]
			\begin{center}

			\includegraphics[scale=0.025]{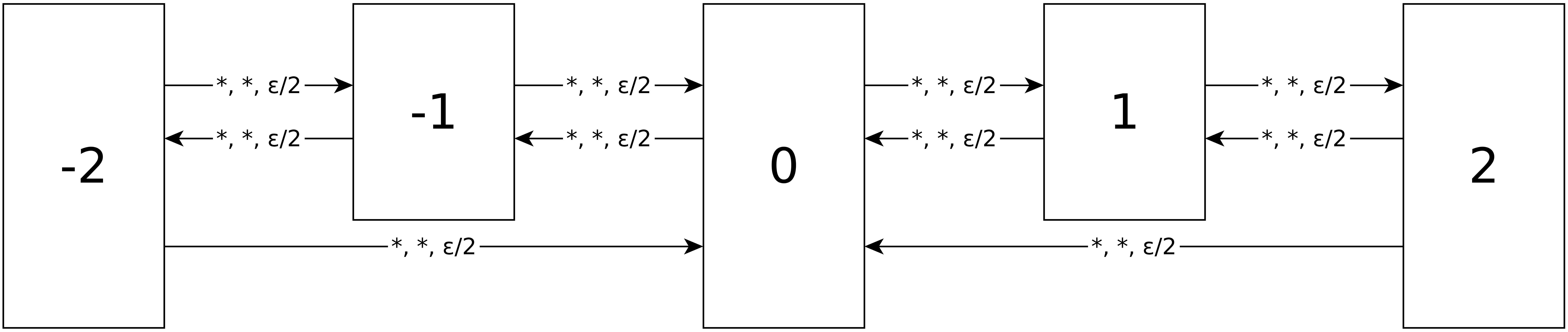}
			
			\caption{Transitions due to network connectivity issues in the repetitive RPS.}
			\label{fig:netrps}
			\end{center}

		\end{figure}
	
\end{example}

\smallskip\noindent{\bf Results about general CMPGs.}
The main results for CMPGs are as follows:
\begin{compactenum}
\item The celebrated result of existence of values was established 
in~\cite{MN81}.

\item For CMPGs, stationary or finite-memory strategies are not 
sufficient for optimality, and even in CMPGs with three states (the well-known 
Big Match game), very complex infinite-memory strategies are required  
for $\epsilon$-optimality~\cite{BF68}. 

\item The value problem, that given a CMPG, a state $s$, and a threshold 
$\lambda$, asks whether the value at state $s$ is at least $\lambda$, can be 
decided in PSPACE~\cite{CMH08}; and also in $m^{2^{O(n)}}$ time, which is doubly 
exponential in the worst case, but polynomial-time in $m$, for $n$ 
constant~\cite{HKLMT11}.
Both the above algorithms use the theory of reals and quantifier 
elimination for analysis. 
\end{compactenum}
\begin{remark}[Inefficiency]\label{rem:inefficiency}
The quantifier elimination approach for general CMPGs considers formulas in the
theory of reals with alternation, where the variables represent  
the transitions~\cite{CMH08}. 
With as few as ten transitions, quantifier elimination produces
formulas with hundreds of variables over the existential theory of reals.
In turn, the existential theory of reals has exponential-time complexity, 
is notoriously hard to solve, and its existing solvers cannot handle hundreds 
of variables. 
Hence, CMPGs with as few as ten transitions are not tractable. 
\end{remark}

\noindent{\bf Results about ergodic CMPGs.}
The main results for ergodic CMPGs, besides the general results for CMPGs, 
are as follows:
\begin{compactenum}
\item Stationary optimal strategies exist\cite{HK66}, 
but positional strategies are not sufficient for optimality.
For precise strategy complexity see~\cite{CI14}.

\item Even in ergodic games, values and probabilities of optimal 
strategies can be irrational~\cite{CI14}, and hence the relevant 
question is the approximation problem of values which is solvable in non-deterministic polynomial-time~\cite{CI14}. 

\item The most well-known algorithm for ergodic mean-payoff games is the 
Hoffman-Karp {\em strategy-iteration} algorithm~\cite{HK66}, which is described in detail in Appendix~\ref{app:hoffman}.

\end{compactenum}
Note that since in ergodic games, every state is reached from every other state
with probability~1, the value at all states is the same.

%% file: modeling.tex
\section{Modeling Framework}\label{sec:Modeling}

In this section we present an abstract framework to model economical 
consequences of attacks with mean-payoff games. In particular 
we show how broad classes of attacks can be modeled as ergodic
games.
In the next section we present concrete examples that arise 
from blockchain protocols. 
We start with some general aspects of mean-payoff games.

\vspace{-0.5em}
\subsection{Mean-payoff games modeling} \label{sec:mean_payoff}

We describe two aspects of mean-payoff games modeling.

\begin{compactenum}
\item {\em Game graph modeling.} 
Graph games are a standard model for reactive systems as well as protocols. 
The states and transitions of the graph represent states and transitions of 
the reactive system, and paths in the graphs represent traces of the 
system~\cite{PR89,RW89}. 
Similarly, in modeling of protocols with different variables for the agents, 
the states of the game represent various scenarios of the protocols along 
with the valuation of the variables.
The transitions represent a change of the scenario along with change in the 
valuation of the variables (for example see~\cite{CR14} for game graph modeling 
of protocols for digital-contract signing). 

\item {\em Mean-payoff objective modeling.} 
In mean-payoff objectives, the costs (or rewards) of every transition
can represent, for example, delays, execution times, cost of context 
switches, cost of concurrency, or monetary gains and losses.
The mean-payoff objective represents the long-term average of the rewards 
or the costs.
The mean-payoff objective has been used for synthesis of better reactive 
systems~\cite{BCHJ09}, synthesis of synchronization primitives for concurrent 
data-structures to minimize average context-switch costs~\cite{CCHRS11}, 
model resource-usage in container analysis and frequency of function calls~\cite{CPV15}, 
as well as analysis of energy-related objectives~\cite{BaierKMNW16,BaierDKKW14,ForejtKP12}.
\end{compactenum}

\subsection{Crypto-currency Protocols as Mean-payoff Games}

We describe how to apply the general framework of CMPGs to crypto-currencies.

\smallskip\noindent\textbf{General setting.}
We propose to analyze protocols as a game between a defender and an attacker.
The defender and the attacker have complete freedom to decide on their moves.
The decisions of the other parties in the ecosystem can be modeled as stochastic choices 
that are not adversarial to either of the players.

\smallskip\noindent\textbf{Reward function.}
The reward function will reflect the monetary gain or loss of the defender.
The attacker gain is not modeled as we consider the worst-case scenario in which 
the attacker's objective is to minimize the defender's utility.

\smallskip\noindent \textbf{States.}
States of the game can represent the information that is relevant for the analysis of the protocol, such as the abstract state of the blockchain.

\smallskip\noindent \textbf{Stochastic transitions.}
Probabilities over the transitions can model true stochastic processes e.g., 
mining, or abstract complicated situations where the exact behavior 
cannot be directly computed 
(see Section~\ref{subsub:double}) or in order to simulate the social behavior 
of a group (see Section~\ref{subsub:bwh}).

\smallskip\noindent\textbf{Concurrent interactions.}
Concurrent games are used when both players need to decide on their action simultaneously or when a single action models a behavior that continues over a time period and the players can only reason about their opponent's behavior after some while (see Sections~\ref{subsub:bwh} and~\ref{subsub:double}).

\smallskip\noindent\textbf{Result of the game.}
In this work we want to reason on defender's security in a protocol 
wrt a malicious attacker who aims to decrease defender's gain at any cost.
The result of the mean-payoff game will describe the inevitable expected loss 
that the defender will have in the presence of an attacker and defender's strategy 
describes the best way to defend himself against such an attacker.

\subsection{Modeling with Ergodic Games}

In this section we describe two classes of attacks, which can be naturally modeled with ergodic games.
Our description here is high-level and informal, and concrete instances are considered in the next 
section. 
The attacks we describe are in a more general setting than crypto-currencies; 
however,
for crypto-currencies the economic consequences are more natural to model.

\smallskip\noindent{\em First class of attacks.}
In the first class of attacks the setting consists of two companies and the revenues of the companies depend 
on the number of users each has.
Thus states represent the number of users. 
Each company can decide to attack its competing company.
Performing an attack entails some economic costs, however it could increase the number of users of the attacking company 
at the expense of the attacked one.
For example, consider two competing social networks, Alice  and Bob.
Alice can decide to launch a distributed-denial-of-service (DDOS) attack on Bob, and vice-versa. 
Such attacks entail a cost, but provide incentives for Bob users to switch to Alice.
The rewards depend on the network revenues (i.e., number of users) and on the amount of funds the company decides 
to spend for the attack.
The migration of users is a stochastic process that is biased towards the stronger network, but with smaller
probability some users migrate to the other network.
Thus the game is ergodic. 
This class represents pool attacks in the context of crypto-currencies (Sections~\ref{subsub:bwh} and~\ref{subsub:pos}).

\smallskip\noindent{\em Second class of attacks.}
Consider the scenario where the state of the game represents aspects of the 
dynamic network topology.
The network evolves over the course of the time, and the actions of the participants also 
affect the network topology.
However, the effect of the actions only makes local changes.
The combination of the global changes and the local effects still ensure that
different network states can be reached, and the game is ergodic.
Attacks in such a scenario where the network topology determines the outcome of 
attack can be modeled as ergodic games.
This class of attacks represent the zero-confirmation double-spending attack in the context of crypto-currencies (see Section~\ref{subsub:double}).

%% file: short_formal_modeling.tex
\vspace{-1em}
\section{Formal Modeling of Real Attacks}\label{sec:formal_short}

In this section we show how to model several real-world examples.
These examples were described in the literature but were never analyzed as stateful games.

\vspace{-1em}
\subsection{Block Withholding Pool Attack}\label{subsub:bwh}

Pools are susceptible to the classic block withholding attack~\cite{rosenfeld2011analysis}, where a miner sends only partial solutions to the pool manager and discards full solutions.
In this section we analyze  block withholding attacks among two pools, pool $A$ and pool $B$.
We describe how pool $A$ can attack pool $B$, and the converse direction is symmetric.
To employ the pool block withholding attack, pool $A$ registers at pool $B$ as a regular miner.
It receives tasks from pool $B$ and transfers them to some of its own miners.
Following the notions in~\cite{eyal2015miner}, we call these infiltrating miners, and their mining
power is called infiltration rate.
When pool $A$'s infiltrating miners deliver partial solutions, pool $A$'s manager submits them to pool $B$'s manager and proves the portion of work they did.
When the infiltrating miners deliver a full solution, the attacking pool manager discards it.

At first, the total revenue of the victim pool does not change (as its effective mining rate was not changed), but the same sum is now divided among more miners.
Thus, since the pool manager fees are nominal (fixed percentage of the total revenue~\cite{managefees}), in the short term, the manager of the victim pool will not lose.
The attacker's mining power is reduced, since some of its miners are used for block withholding, but it earns additional revenue through its infiltration of the other pool.
Finally, the total effective mining power in the system is reduced, causing the blockchain
protocol to reduce the difficulty. Hence, in some scenarios, the attacker can gain, even in the short run, from performing the attack~\cite{eyal2015miner}.

In the long run, if miners see a decrease in their profits (since they have to split the same revenue among more participants), it is likely that they consider to migrate to other pools.
As a result, the victim pool's total revenue will decrease.

\smallskip\noindent \textbf{Our modeling.}
We aim to capture the long term consequences of pool attacks.
We have two pools $A$ and $B$, where $B$ is the victim pool and $A$ is the malicious pool who wishes to decrease $B$'s profits.
There is also a group of miners $C$ who are honest and represent the rest of the network.
In return, pool $B$ can defend itself by attacking back.
To simulate the long term effect, in every round pool members from $A$ and $B$ may migrate from one pool to another or to and from $C$.
The migration is a stochastic process that favors the pool with maximum profitability for miners.
We note that given sufficient amount of time (say a week), a pool manager can evaluate with very high probability the fraction of infiltrating miners in his pool.
This can be done by looking at the ratio between full and partial solutions.
Hence, in retrospect of a week, the pools are aware of each other's decisions, but within this week there is uncertainty. Therefore, we use concurrent games to analyze the worst case scenario for pool $B$.

\begin{theorem}\label{theo:bw}
Consider a pair of pools $A$ and $B$ capable of attacking each other. Let $C$ be the pool of remaining miners. If the miners in each pool migrate stochastically according to the attractiveness levels (as detailed below), then $B$ can ensure a revenue of at least $v$ on average per round, against any behavior of $A$, where $v$ is the value of the concurrent ergodic game described below.
\end{theorem}

\subsubsection{Details of Modeling}

We provide details of our modeling to demonstrate how such attacks can be thought of in terms of ergodic games. Due to page limitation and similarity, such details in other cases are relegated to Appendix~\ref{sec:appen_formal_modeling}.
\begin{compactitem}
	\item {\em Game states.} We consider two pools, $A$ and $B$ and assume that any miner outside these two is mining independently for himself. Each state is defined by two values, i.e. the fractions of total computation power that belongs to $A$ and $B$. We use a discretized version of this idea to model the game in a finite number of states and let $S = \{1, 2, \ldots, n\}^2$ and define $\epsilon = \frac{1}{2n+1}$, where a state $(i_1, i_2) \in S$ corresponds to the case where pool $A$ owns a fraction $\alpha_{i_1} = i_1 \epsilon = \frac{i_1}{2n+1}$ of the total hash power and pool $B$ controls a fraction $\beta_{i_2} = i_2 \epsilon =  \frac{i_2}{2n+1}$ of it. In this case the miners who work independently own a fraction $\gamma_{i_1,i_2} = 1 - \alpha_{i_1} - \beta_{i_2}$ of the total hash power.
	\item {\em Actions at each state.}
	Each pool can choose how much of its hash power it devotes to attacking the other pool. More formally, at each state $s = (i_1, i_2)$, pool $A$ has $i_1$ choices of actions and $\Gamma_1(s) = \{a_1^0, a_1^1, a_1^2, \ldots, a_1^{i_1 - 1} \}$ where $a_1^j$ corresponds to attacking pool $B$ with a fraction $j \epsilon$ of the total computing power of the network. Similarly $\Gamma_2(s) = \{a_2^0, a_2^1, a_2^2, \ldots, a_2^{i_2-1}\}$.
	
	\item {\em Rewards.} We want the rewards to model the revenue (profit) of pool $A$, denoted by $r_A$, so we let 
	$R(s, a_1^i, a_2^j) = r_A (s, a_1^i, a_2^j),$
	for $a_1 \in \Gamma_1(s), a_2 \in \Gamma_2(s)$.
	We write $r_A$ instead of $r_A(s, a_1^i, a_2^j)$ when there is no risk of confusion. We define $r_B$ and $r_C$ similarly and normalize the revenues: $r_A + r_B + r_C = 1$.
	
	To compute these values, we define ``attractiveness''.
	The attractiveness of a pool is its revenue divided by the total computing power of its miners.
	
	If pool $A$ chooses the action $a_1^i$ and pool $B$ chooses the action $a_2^j$, then pool $A$ is using a fraction $\alpha' = i \epsilon$ of the total network computing power to attack $B$ and is receiving a corresponding fraction of $B$'s revenue while not contributing to it. Therefore the attractiveness of pool $B$ will be equal to: 
	$
	attr_B = \frac{r_B}{\beta + \alpha'}.
	$
	Similarly we have
	$
	attr_A = \frac{r_A}{\alpha + \beta'},
	$
	where $\beta' = j \epsilon$.
	
	Now consider the sources for pool $A$'s revenue. It either comes from $A$'s own mining process or from collecting shares of $B$'s revenue, therefore:
	\[
	r_A = (\alpha - \alpha') + \alpha' \times attr_B,
	\]
	and similarly
	$
	r_B = (\beta - \beta') + \beta' \times attr_A.
	$
	The previous four equations provide us with a system of linear equations which we can solve to obtain the values of $r_A$, $r_B$, $attr_A$ and $attr_B$.
 Since a fraction $\alpha' + \beta'$ of total computation power is used on attacking other pools, we have:
	$
	attr_C = \frac{1}{1 - \alpha' - \beta'}.
	$
	
	\item {\em Game transitions ($\delta$).}
	Miners migrate between pools and a pool gains or loses mining power based on its attractiveness. If a pool is the most attractive option among the two, it gains $\epsilon$ new mining power with probability $\frac{2}{3}$, retains its current power with probability $\frac{1}{6}$ and loses $\epsilon$ power with probability $\frac{1}{6}$. 
	On the other hand a pool that is not the most attractive option loses $\epsilon$ power with probability $\frac{2}{3}$, retains its current power with probability $\frac{1}{6}$ and attracts $\epsilon$ new mining power with probability $\frac{{1}}{6}$. These values were chosen for the purpose of demonstration of our algorithm and our implementation results. In practice, one can obtain realistic probabilities experimentally.

	\item {\em Ergodicity.} The game is ergodic because for each two states $s = (s_1, s_2)$ and $s' = (s_1', s_2')$ where $\vert s_1 - s_1' \vert \leq 1$ and $\vert s_2 - s_2' \vert \leq 1$, there is at least $\frac{1}{36}$ probability of going from $s$ to $s'$ no matter what choices the players make.
	
\end{compactitem}

\noindent \textbf{Proof of Theorem \ref{theo:bw}.}
Ergodicity was established in the final part above. The rest follows from the modeling and the determinacy result.

\vspace{-0.6em}
\subsection{Zero-confirmation Double-spending}\label{subsub:double}
\vspace{-0.3em}

Nowadays, Bitcoin is increasingly used in ``fast payments'' such as online services, ATM withdrawals and vending machines~\cite{cnn},
where the payment is followed by fast delivery of goods.
While the blockchain consensus is appropriate for slow payments, it requires tens of minutes to confirm a transaction
and is therefore inappropriate for fast payments. We consider a transaction confirmed when it is added to the blockchain and several blocks are added after it.
This mechanism is essential for the detection of double-spending attacks in which an adversary attempts to use some of her coins for two or more payments.
However, even in the absence of a confirmation, it is far from trivial to perform a double-spending attack.
In a double spending attack, the attacker publishes two transactions that consume the same input.
The attack is successful only if the victim node received one transaction and provided the goods before he became aware of the other, but eventually the latter was added to the blockchain.
In an ideal world the attacker can increase his odds by broadcasting one transaction directly to the victim and the other at a far apart location, while on the other hand the victim can defend itself by deploying several nodes in the network in \emph{strategic} locations.
In the real world, however, the full topology of the network is never known to either of the parties.
Nevertheless, based on history and network statistics one can estimate the odds of a successful attack given the current state of the network~\cite{blockchyper}.

The victim has to decide on a policy for accepting zero-confirmation transactions.
In particular he has to decide on the probability of whether to wait for a confirmation or not.
If he waits for confirmation, then the payment is guaranteed, but customer satisfaction is damaged, and as a result the utility is smaller than the actual payment.
If he does not wait for a confirmation, then the payment might be double spent. 
In the long term, the victim could decide to change the topology of the network.
As it does not have full control over the topology, the outcome of the change is stochastic.
Moreover, even when the victim does not initiate a change, the network topology is dynamic and keeps changing all the time.
Hence, the odds of a successful attack are constantly changing in small stochastic steps.

\smallskip\noindent \textbf{Our modeling.}
We aim to analyze the worst case long run loss of the victim.
In our model we abstract the network topology state and consider only the odds of successful double spending.
We consider a scenario where the victim's honest customers typically purchase goods worth 10 units per round.
In every round, the victim decides on a policy for accepting fast payment, and the attacker, concurrently, unaware of the victim's policy, has to decide the size of the attack.
After every round, the victim decides if he wants to do a thorough change in the network topology.
If he decides on a change, then the next state is chosen uniformly from all possible states (this represents the fact that neither players has full knowledge on the topology).
If he decides to make no change, then the network state might still change, due to the dynamic nature of the network.
In this case the next state is with high probability either the current state, or a state which is slightly better or slightly worse for the victim, but with low probability the state changes completely to an arbitrary state in the network (as sometimes small changes in the topology have big impact).
The rewards stem from the outcome of each round in the following way:
The payment is the sum of the honest customer purchases and the payment of the attacker (if it gets into the blockchain).
The reward is the payment minus some penalty in case the victim has decided to wait for a confirmation.
The fact that the network state is constantly changing makes our model ergodic.
A proof and more details of the following Theorem are provided in Appendix~\ref{app:model_zc}.
\begin{theorem}\label{theo:ds}
Consider a seller and an attacker in the zero-confirmation double spending problem. The seller can ensure profit of at least $v$ on average per round, where $v$ is the value of the corresponding CMPG.
\end{theorem}

\subsection{Proof of Stake Pool Attack}\label{subsub:pos}

Proof of stake protocols allow miners to centralize their stakes in a pool.
In such pools the withholding attack is not relevant as mining does not require any physical resources.
However, pool $A$ might attack an opponent pool $B$ by not signing or broadcasting its blocks.
A successful attack would prevent the block from getting signed by a majority of the network.
The result would be a loss of mining fees for $B$ and can encourage miners to migrate from the pool.
An unsuccessful attack decreases $A$'s signing fee revenue.

\smallskip\noindent \textbf{Our modeling.}
We assume a setting similar to that of Section~\ref{subsub:bwh}, where there are two opponent pools $A$ and $B$, and the rest of the network consists of honest pools who sign every block that arrives on time.
The states of the game are the stakes of each pool, namely $\alpha$ for pool $A$ and $\beta$ for pool $B$.
In every round, with probability $1-(\alpha+\beta)$ neither of the pools is elected to mine a block, and no decisions are made.
Otherwise, with probability $\frac{\alpha}{\alpha+\beta}$ pool $A$ is elected and otherwise pool $B$ is elected.
When a pool is elected, the other pool decides whether to sign and broadcast the resulting block or not.
In addition the network state and connectivity induce a distribution over the fraction of honest miners that receive the block.
If the block is accepted, then its creator is rewarded with mining fees, and the other pool will get its signing fees only if it signed the block.
A proof and more details of the following Theorem are provided in Appendix~\ref{app:model_stake}

\begin{theorem}\label{theo:ps}
Consider two pools $A$ and $B$ in a proof of stake mining system that can choose to attack each other by not signing blocks mined by the other pool. 
Consider that the rest of the network consists of independent miners who observe published blocks according to a predefined probability distribution and sign every valid block they observe. If the miners migrate according to the attractiveness levels (as described in Section \ref{subsub:bwh}), then $B$ can ensure an average revenue of $v$ against any behavior of $A$, where $v$ is the value of the corresponding CMPG.
\end{theorem}

%% file: experimental_results.tex
\section{Implementation and Experimental Results}

In this section we present our implementation details and experimental results. The code is available at \url{http://ist.ac.at/~akafshda/concur2018}.

\input{implementation}

\vspace{-1.0em}
\subsection{Experimental Results} 
\vspace{-0.5em}

We provide experimental results for all games in Section~\ref{sec:formal_short}. 
We show  number of transitions in the game (\#T), number of states in the game, the running time and number of strategy iterations (\#SI) for every scenario.

\begin{table}[!ht]
	\vspace{-2mm}
	\begin{center}
	\begin{minipage}{0.32\linewidth}
		\resizebox{\textwidth}{!}{
		\begin{tabular}{ |c | c | c | c| }
			\hline
			\#T & States & \#SI & Time(s) \\
			\hline \hline
			17050 & 100 &  4 & 69\\
			56252 & 196 &  2 & 291\\
			135252 & 289 &  2 & 389\\
			236000 & 400 & 2 & 1059\\
			331816 & 484 & 2 & 3880\\
			508032 & 576 &  2 & 6273\\
			720954 & 676 &  2 & 17014\\
			966281 & 784 & 2 & 53103\\
			1269450 & 900 & 2 & 100435\\
			\hline
		\end{tabular}}
	\end{minipage}
	\begin{minipage}{0.32\linewidth}
		\resizebox{\textwidth}{!}{
		\begin{tabular}{ |c | c | c | c|}
			\hline
			\#T & States &  \#SI & Time(s) \\
			\hline \hline
			19940 & 100 &  2 & 426 \\
			40040 & 200 &  2 & 800 \\
			60140 & 300 & 2 & 1141 \\
			80240 & 400 &  2 & 1586\\
			100340 & 500 & 2 & 2069\\
			120440 & 600 &  2 & 1253\\
			140540 & 700 & 2 & 2999\\
			160640 & 800 & 2 & 3496\\
			180740 & 900 & 2 & 3917\\
			\hline
		\end{tabular}}
	\end{minipage}
	\begin{minipage}{0.32\linewidth}
		\resizebox{\textwidth}{!}{
		\begin{tabular}{ |c | c | c | c| }
			\hline
			\#T & States & \#SI & Time(s) \\
			\hline \hline
			6076 & 99 &  18 & 471\\
			20956 & 275 &  8 & 1338\\
			31744 & 396 &  9 & 2520\\
			44764 & 539 &  4 & 1073\\
			77500 & 891 &  16 & 22125\\
			119164 & 1331 &  27 & 32636\\
			169756 & 1859 &  10 & 31597\\
			262384 & 2816 &  12 & 89599\\
			\hline
		\end{tabular}}
	\end{minipage}
	\vspace{1mm}
	\caption{Experimental results for block-withholding pool attack (left), zero-confirmation double-spending (center) and proof of stake pool attack (right).}
	\label{table:experimental_results}
\vspace{-10mm}
\end{center}
\end{table}

Note that \#SI is not monotone in the number of states.
Intuitively the number of needed iterations depends on the extent in which easy locally optimal strategies are also globally optimal.
In addition the strategy iteration algorithm starts with an arbitrary random strategy, and hence the number of iterations also depends on the initial strategy. However, it is worthy to note that in all cases the number of iterations required is quite small.  We also note that since the number of iterations is small, the crucial computational step is every iteration, where many linear-programming problems are solved. 

\smallskip\noindent \textbf{Outputs of the algorithm.} The outputs provided the following results:
\begin{compactitem}
\item For the block withholding pool attack game, the algorithm could guarantee a mean-payoff of $0.49$ for the victim pool. In absence of an attacker the pool becomes the most attractive option for miners and grows to maximum possible size with probability $1$, hence if there is no other pool the mean-payoff will be $1$. Also, if there are two pools $A$ and $B$ with hash powers $\alpha$ and $\beta$ respectively, and they decide not to attack each other, then they will both become the most attractive option and will grow with the same rate, leading to a mean-payoff of $\alpha + \frac{1 - \alpha - \beta}{2}$ for $A$ and $\beta + \frac{1-\alpha-\beta}{2}$ for $B$.
\item For the zero-confirmation double-spending game, the algorithm verified that the seller is guaranteed to maintain at least half of her revenue, i.e., in presence of a malicious attacker,
the value for the seller converges to $5$ as the number of states increase, while it is $10$ in absence of it.
\item For the proof of stake pool attack game, by increasing the number of states, i.e., by refining the discretization, the guaranteed value (game value) decreases and tends to zero. In absence of an attacker, a pool $A$ can achieve an expected payoff of $11 s_A$ at a turn where $s_A$ is the stake it holds. This is because it earns an average of $10 s_A$ from mining fees and $s_A$ from signing. In this case, since the pool becomes the most attractive option, it gains miners and reaches a stake of $1$, leading to a mean-payoff of $11$.
\end{compactitem}

For the exact details see Appendix~\ref{sec:appen_formal_modeling}. Our algorithm also finds strategies that achieve these values. 

%% file: implementation.tex

\subsection{Implementation Challenges}

We have implemented the strategy-iteration algorithm for ergodic games (see Appendix~\ref{app:hoffman} for pseudo-code and more details).
To the best of our knowledge, this is the first implementation of this algorithm.
The straightforward implementation of the strategy-iteration algorithm for
ergodic games has two practical problems, which we describe below.
\begin{compactenum}
\item {\em No stopping criteria.} 
First, the strategy-iteration algorithm only guarantees convergence of values 
in the limit, and since values and probabilities in strategies can be 
irrational, convergence cannot be guaranteed in a finite number of steps. 
Hence we need a stopping criterion for approximation.

\item {\em Numerical precision issues.} Second, the stationary strategies in each iteration 
are obtained through solution of linear-programming, which has numerical errors, and the probabilities
sum to less than~1. 
If these errors remain, they cascade over iterations, and do not ensure convergence in practice
for large examples.
Hence we need to ensure numerical precision on top of the strategy-iteration algorithm.
\end{compactenum}
Our solution for the above two problems are as follows:
\begin{compactenum}
\item {\em Stopping criteria.} 
We first observe that the value sequence which is obtained converges from below to the 
value of the game. 
In other words, the value sequence provide a lower bound to the lower value of the game.
Hence we consider a symmetric version which is the strategy-iteration algorithm for player~2,
and run each iteration of the two algorithms in sequence. 
The version for player~2 provides a lower bound on the lower value for player~2, and thus from
that we can obtain an upper bound on the upper value of player~1.
Since the upper and lower values coincide, we thus have both an upper and lower bound on the 
values, and once the difference is smaller than $\epsilon>0$, then the algorithm has correctly
approximated the value within $\epsilon$ and can stop and return the value and the strategy 
obtained as approximation.

\item {\em Numerical precision.} 
For numerical precision, instead of obtaining the results from the linear program, we 
obtain from the linear program the set of {\em tight} and {\em slack} constraints, where
the tight constraints represent the constraints where equality is obtained, and the other
constraints are slack ones.
From the tight constraints, which are equalities, we obtain the result using Gaussian elimination,
which provides more precise values to the solution. 
We also provide other heuristics, such as adding the remaining probability to the greatest
probability action, and obtain similar results on convergence. 
\end{compactenum}


%% file: related.tex
\vspace{-1em}
\section{Related Work}\label{sec:related}

\noindent{\em Basic bitcoin security.}
The first security analysis of the Bitcoin protocol was done by Nakamoto~\cite{nakamoto2008bitcoin} who showed the resilience of the blockchain protocol against a double-spending attack.
His analysis was later corrected by Rosenfeld~\cite{Rosenfeld14} who showed that the use of probabilistic arguments in the original analysis was not sound.
Rosenfeld's analysis gives different numerical results, but still certifies the original security properties.
Recently Sompolinsky and Zohar~\cite{SompolinskyZ16} further refined the analysis by considering the fact that the attacker can observe the possible states of the blockchain before choosing to attack, and thus he can increase his utility by choosing the right time to attack.

\smallskip\noindent{\em Pools attack.}
The danger of a block withholding attack is as old as Bitcoin
pools. The attack was described by Rosenfeld~\cite{rosenfeld2011analysis} as early
as 2011, as pools were becoming a dominant player in the
Bitcoin world.
While it was obvious that a pool is vulnerable to a malicious attacker,
Eyal~\cite{eyal2015miner} showed that in some circumstances a pool can benefit by attacking another pool, and thus pool mining is vulnerable also in the presence of rational attackers.
However, the analysis only considered the short term, i.e., the profit that the pool can get only in the short period after the attack.
Laszka et al.~\cite{laszka2015bitcoin} studied the long term impact of pools attack.
In their framework miners are allowed to migrate from one pool to another.
They analyzed the steady equilibrium in which the size of the pools become stable (although there is no guarantee that the game will converge to such a scenario).
Our framework is the first to allow analysis of long term impacts without convergence assumptions.

\smallskip\noindent{\em Zero-confirmation double-spending.}
Zero-confirmation double-spending was experimentally analyzed by Karame et al.~\cite{karame2012two} who gave numerical figures for the odds of successful double spending for different network states.
However, their analysis did not consider the fact that the victim may change his connectivity state.
Our work is the first analysis framework for the long term impact of zero-confirmation double-spending.

\smallskip\noindent{\em Stateful analysis.}
A stateful analysis of blockchain attacks was done by Sapirshtein et al.~\cite{sapirshtein2015optimal} and by Sompolinsky and Zohar~\cite{SompolinskyZ16}.
In their analysis the different states of the blockchain were taken into account during the attack.
The analysis was done using MDPs (a single player game, where only one player makes the choices) in which only the attacker decides on his actions and the victim follows a predefined protocol.
A recent work \cite{smartcontractpaper} has also considered abstraction-refinement for finite-horizon games in the context of smart contracts. However, it neither considers long-term behavior, nor mean-payoff objectives, nor can it model attacks such as double-spending and interactions between pools (see Appendix~\ref{app:comparison} for more details).

\smallskip\noindent{\em Quantitative verification with mean-payoff games.}
The mean-payoff games problem has been studied extensively as a theoretical
problem in various models \cite{PR89,RW89}. 
The mean-payoff games problem has also been studied in the context of 
verification and synthesis for performance related issues \cite{BCHJ09,CCHRS11,CPV15,BaierKMNW16,BaierDKKW14,ForejtKP12} 
(see Section~\ref{sec:mean_payoff} for more details).
However all these works focus on turn-based games, and none of them 
consider concurrent games.
To the best of our knowledge concurrent mean-payoff games have 
not been studied in the setting of security that we consider, where 
the quantitative objective is as crucial as safety critical issues.
Practical implementation of algorithms for ergodic CMPGs 
do not exist in the literature.

\smallskip\noindent{\em Formal methods in security.}
There is a huge body of work on program analysis for security
(see~\cite{SabelfeldM03,Abadi12} for surveys).
Formal methods are used to create safe programming languages (e.g.,~\cite{fuchs2009scandroid,ZhangWSM15,SabelfeldM03}) and to define new logics that can express security properties (e.g.,~\cite{burrows1989logic,flac,ArdenLM15}).
They are also used to automatically verify security and cryptographic protocols, e.g.,~\cite{abadi2000reconciling,blanchet2008automated} and~\cite{avalle2014formal} for a survey.
However, all of these works aimed to formalize qualitative properties such as privacy violation and information leakage.
The works of~\cite{KremerR03,CR14} consider analysis of security protocols with turn-based games and qualitative properties.
To our knowledge, our framework is the first attempt to use
concurrent mean-payoff games as a tool for reasoning about economic effects
of attacks in crypto-currencies.

%% file: conclusion.tex
\vspace{-0.5em}
\section{Conclusion and Future Work}\label{sec:conculsion}

In this work we considered concurrent mean-payoff games, and in particular 
the subclass of ergodic games, to analyze attacks on crypto-currencies.
There are several interesting directions to pursue:
First, various notions of rationality are relevant to analyze games where the attacker is rational, 
rather than malicious, and aims to  maximize his own utility instead of minimizing the defender's utility
(e.g., secure-equilibria~\cite{ChatterjeeHJ04} or other related notions). 
Second, we consider two-player games, and the extension to multi-player games to model
crypto-currency attacks is another interesting problem.

%% file: appendix.tex
\appendix

\section{The Hoffman-Karp Strategy-iteration Algorithm}  \label{app:hoffman}

For an ergodic CMPG $G$ and a state $t$, the basic informal description of 
the algorithm is as follows.
In every iteration $i$, the algorithm considers a stationary strategy 
$\stra_1^i$, and then improves the strategy locally as follows: 
first it computes the \emph{potential} $v^{\stra_1^i}_s$ (described below) 
given $\stra_1^i$, and then for every state $s$, the algorithm locally 
computes an arbitrary optimal distribution at $s$ to improve the potential.
The intuitive description of the potential is as follows: 
Fix the specific state $t$ as the target state (where the potential must be~0); 
and given a stationary strategy $\stra$, consider a modified reward function 
that assigns the original reward minus the value ensured by $\stra$.
Then the potential for every state $s$ other than the specified state $t$ is 
the expected sum of rewards under the modified reward function for the random 
walk from $s$ to $t$.
The local improvement step is obtained as a solution of a matrix game with 
the potentials.
The formal description of the algorithm is given in Figure~\ref{alg-CMPG},
and the formal definition of the expected one-step reward $\ExpRew$ and 
one-step function $\OneStep$ is below.

\smallskip\noindent{\em Notations: $\ExpRew$ and $\OneStep$.}
The expected one-step reward $\ExpRew(s,\stra_1^{i},a_2)$ for a stationary strategy 
$\stra_1^{i}$ for Player~1, that specifies a distribution $\stra_1^{i}(s)$ for every
state, and an action $a_2 \in \mov_2(s)$, is as follows:
\[
\ExpRew(s,\stra_1^{i},a_2)=\sum_{a_1 \in \mov_1(s)} 
\cost(s,a_1,a_2) \cdot \stra_1^{i}(s)(a_1) \enspace .
\]
Similarly, we will also use the following notation:
\[
\trans(s,\stra_1^{i},a_2)(s')= 
\sum_{a_1 \in \mov_1(s)} \trans(s,a_1,a_2)(s') \cdot \stra_1^{i}(s)(a_1) 
\enspace .
\]
For notional convenience, given a vector $x=(x_i)_{i\in S}$, a state $s$ and a pair of distributions $d_1\in\distr(\mov_1(s))$ and $d_2\in \distr(\mov_2(s))$, we let $\OneStep(x,d_1,d_2,s)$ be\[
\OneStep(x,d_1,d_2,s)=\sum_{\substack{a_1\in\mov_1(s)\\a_2\in\mov_1(s)\\s'\in S}}d_1(a_1)\cdot d_2(a_2)\cdot\trans(s,a_1,a_2)(s') \cdot x_{s'}\enspace .
\]
Also, given a vector $x=(x_i)_{i\in S}$, a state $s$ and a  stationary strategy profile $\stra=(\stra_1,\stra_2)$, we will let $\OneStep(x,\stra,s)$ be 
\[
\OneStep(x,\stra,s)=\OneStep(x,\stra_1(s),\stra_2(s),s)
\]

\begin{figure}
	\begin{center}
		{\footnotesize
			\begin{function}[H]
				Let $\stra_1^1$ be a Player-1 stationary strategy\;
				\For{$(i\in \Z_+)$}{
					Compute $g^i,(v_s^i)_{s\in S}$ as the unique solution of 
					\begin{align*}
					g^i+v^i_s & = \min_{a_2\in \mov_2(s)}(\ExpRew(s,\stra_1^{i},a_2)+
					\OneStep(v^i,\stra_1^{i}(s),a_2,s) \\
					& \qquad \quad \forall  s\in S \\ 
					v^i_t & =0;
					\end{align*}
					\For{$(s\in S)$}
					{
						Let $M_s$ be the matrix game defined as follows\;
						\For{($a_1\in \mov_1(s)$ and $a_2\in \mov_2(s)$)}
						{
							$P_s[a_1,a_2]:=  \OneStep(v^i,a_1,a_2,s)$\;
							$M_s[a_1,a_2]:= \cost(s,a_1,a_2) + P_s[a_1,a_2]$\; 
						} 
						\If{{\em $(\stra_1^{i}(s)$ is an optimal distribution for $M_s)$}}
						{
							$\stra_1^{i+1}(s) := \stra^{i}_1(s)$\;
						}
						\Else
						{
							$\stra_1^{i+1}(s):=$ Optimal distribution over $\mov_1(s)$ for $M_s$\;
						}
					}
					\If{$(\stra_1^{i+1}= \stra_1^{i})$}{\Return{$\stra_1^i$}\;}
				}
				\caption{HoffmanKarp($G$,$t$)}
			\end{function}
		}
	\end{center}
	\vspace*{-0.5cm}
	\caption{\label{alg-CMPG} Strategy-iteration algorithm for solving ergodic games}
	
\end{figure}

\smallskip\noindent{\bf Computation of every iteration.} 
The computation of every iteration is as follows. 
The computation of the unique solution $g^i$ and $(v^i_s)_{s\in S}$ is 
obtained in polynomial time using linear programming. 
The fact that the solution is unique follows from the fact that once a 
strategy for Player~1 is fixed, we obtain an MDP for Player~$2$, and 
then the MDP solution is unique.
The value sequence $(g^i)_{i\geq 1}$ obtained by the Hoffman-Karp algorithm 
converges to the value $g$ of the game~\cite{HK66}.

\section{Comparison with other game-theoretic works} \label{app:comparison}
Previous works~\cite{smartcontractpaper,bonneau2015sok} consider either one-shot or finite-horizon games for security analysis. 
In contrast, the main differences of our work are as follows:
\begin{itemize}
	\item Finite-horizon (or bounded-horizon) games can be reduced to one-shot games with an exponential blow up in the number of strategies. 
	Thus one-shot and finite-horizon games are conceptually similar, though there are computational complexity differences. 
	In contrast to finite-horizon games, we consider infinite-horizon games, which is conceptually different from finite-horizon games and there is no reduction (even with a blow up) to finite-horizon or one-shot games.
	\item In finite-horizon games for crypocurrency, the focus is on abstraction-refinement~\cite{smartcontractpaper}. 
	In contrast, we consider a special class of games (ergodic games) and use algorithmic approaches for finding their values.
	\item In this work we consider attacks that are inherent in the Blockchain, such as double-spending and pool-attacks. Previous works do not consider the analysis of such attacks. 
\end{itemize}

%% file: formal-modeling.tex
\section{Formal Modeling of Problems as Concurrent Games}\label{sec:appen_formal_modeling}

\subsection{Formal Modeling of Zero-confirmation Double-spending}  \label{app:model_zc}

\begin{itemize}
	
	\item {\em Game states}.
	Each state of the game corresponds to a probability of success for double spending attack which is an abstraction of the network topology.
	
	We discretized the game into $n+1$ states and set \[S = \{0, 1, \ldots, n\}.\] The state $0$ is called a shuffling state. Each other state $i$ corresponds to a double spending success probability of $p_i = 0.1 + \frac{(i-1) \times 0.4}{n}$ for $1 \leq i \leq n$. Player 1 is the seller and player 2 is the malicious buyer.
	
	\item{\em Actions at each state.}	
	The shuffling state, 0, corresponds to the seller deciding to disconnect and reconnect to the network so as to randomly obtain one of the other states. Therefore each player has only one action, i.e. no choice, in this state. We denote these actions by $a_1^0$ and $a_2^0$ respectively.
	
	At each other state the seller can decide to disconnect from the network and then reconnect to it. Moreover he can choose whether to require a confirmation and wait for it, so the seller has $4$ possible actions. We denote these actions by $a_1^{i}$ as in the following table:
	
	\begin{center}
		\begin{tabular}{ | c | c | c| }
			\hline
			action & reconnect & confirmation \\ \hline \hline 
			$a_1^0$ & No & No \\ \hline
			$a_1^1$ & Yes & No\\ \hline
			$a_1^2$ & No & Yes\\ \hline
			$a_1^3$ & Yes & Yes\\ \hline
		\end{tabular}
	\end{center}
	
	The malicious buyer can decide how much double spending to attempt. He can attempt between $1$ and $20$ units of double spending. We denote the action of attempting $d$ units of double spending as $a_2^d$.
	
	\item {\em Game transitions.}
	If the game is in the shuffling state, the next state will be one of the other $n$ states and all of them are equally likely, i.e.
	\[
	\delta(0, a_1^0, a_2^0)(i) = \frac{1}{n},
	\] 
	for $1 \leq i \leq n$.
	
	Otherwise, if the seller decides to reset his connection to the network, the game will be transitioned to the shuffling state with probability 1, no matter what choice was made by the buyer. More formally,
	\[
	\delta(s, a_1^1, a_2)(0) = \delta(s, a_1^3, a_2)(0) = 1,
	\] 
	for all $s \in S \setminus \{0\}$ and $a_2 \in \Gamma_2(s)$.
	
	Otherwise, if the seller decides to wait for a confirmation, then the attack will be unsuccessful and if he does not wait for a confirmation, then the current state defines the odds of a successful attack. So the attack will succeed with probability 
	\[
	p_a(s, a_1) =  \left\{\begin{matrix}
	p_s & a_1 \in \{a_1^0, a_1^1\} \\ 
	0 & a_1 \in \{a_1^2, a_1^3\} 
	\end{matrix}\right. .
	\]
	
	We consider two cases. If the attack is successful, the game transitions to state $n$, i.e. the state with the highest probability of double-spending success, 
	\[
	\delta_a(s, a_1, a_2)(n) = p_a(s, a_1),
	\]
	for $s \in S \setminus \{0\}, a_1 \in \{a_1^0, a_1^2\}$ and $a_2 \in \Gamma_2(s).$
	Intuitively, this is because if the attacker was successful once, he can repeat the attack. 
	
	If the attack fails, then if the game is at state $s$, it goes to each of the states $s-1, s$ and $s+1$ (if they exist and are non-zero) with equal probability. This captures small changes in the topology of the network that can be caused by factors that are not parties to the game, like other people reconnecting to the network. Also the game will transition to the shuffling state with a small probability $p_{dc}$. This models the natural loss of connection that may occur in the network and cause the seller to reconnect even though he did not intentionally decide to do so. In the implementation we set $p_{dc} = 0.001$.  More formally by letting $N_s = \{s-1, s, s+1\} \cap \{1, 2, \ldots, n\}$, we have
	\[
	\delta_b(s, a_1, a_2) (0) = p_{dc} (1 - p_a (s, a_1)),
	\]
	\[
	\delta_b(s, a_1, a_2) (s') = \frac{1-p_a(s, a_1)}{\vert N_s \vert} \times (1 - p_{dc}),
	\]
	where $s \in S \setminus \{0\}$, $s' \in N_s$, $a_1 \in \{a_1^0, a_1^2\}$ and $a_2 \in \Gamma_2(s)$.
	
	Finally we set the probability of transitioning to a state $s'$ as the sum of the two probabilities obtained in the cases above:
	\[
	\delta(s, a_1, a_2)(s') = \delta_a(s, a_1, a_2)(s') + \delta_b(s, a_1, a_2)(s')
	\]
	for $s, s' \in S \setminus \{0\}$, $a_1 \in \{a_1^0, a_1^2\}$ and $a_2 \in \Gamma_2(s)$.
	
	\item {\em Rewards.} The rewards model net income (profit) of the seller.
	Transitions from the shuffling state carry a reward of zero, since the seller is unable to sell any goods while his connection is being reset. Formally,
	\[
	R(0, a_1^0, a_2^0) = 0.
	\]
	
	Assuming that the seller has a profit ratio of $p$. We model the rewards to capture his profit. In the implementation we have set $p = 0.5$. Recall that the malicious buyer, when choosing action $a_2^d$,  is trying to double-spend an amount $d$, $1 \leq d \leq 20$ and that other buyers are interested in buying $10$ units of goods from the seller.
	
	Again we consider two cases. If the double spending attack is successful, this will yield to a total payoff of $-d (1-p)$ for the seller while an unsuccessful attack gives him a profit of $dp$. So we can set
	\[ R_1(s, a_1, a_2^d) = d p (1 - p_a (s, a_1))- d (1-p) p_a (s, a_1) \]
	for $s \in S \setminus \{0\}, a_1 \in \Gamma_1(s)$ and $1 \leq d \leq 20$.
	
	Now we focus on the profit of selling to other (non-malicious) buyers. If the seller decides to wait for a confirmation, he will not be able to serve a fraction $f$ of his other customers, who are not willing to wait. We have set $f=0.5$ in the implementation. So he gets a total revenue of $10 p (1 - f)$ from his other customers. On the other hand, if he does not wait for a confirmation he will receive a payoff of $10p$, so 
	\[
	R_2(s,a_1^0 ,a_2^d) = R_2(S, a_1^1, a_2^d) = 10 p,
	\]
	\[
	R_2(s, a_1^2, a_2^d) = R_2(s, a_1^3, a_2^d) = 10 p (1-f),
	\] 
	for $s \in S \setminus \{0\}$ and $1 \leq d \leq 20$.
	
	The final payoff is the sum of profits that the seller makes by selling to the malicious buyer and others, i.e.
	\[
	R(s, a_1, a_2) = R_1(s, a_1, a_2) + R_2(s, a_1, a_2)
	\]
	for $s \in S \setminus \{0\}$, $a_1 \in \Gamma_1(s)$ and $a_2 \in \Gamma_2(s)$. 
	
	\item {\em Ergodicity.} This game is ergodic. Starting with any state and strategy profile, the shuffling state, 0, is visited infinitely often with probability 1. This is because any choice of actions by the two players at each turn would switch the game to the shuffling state with probability at least $p_{dc}$. Since the shuffling state is visited infinitely often, and since all other states have a non-zero probability of following the shuffling state, we conclude that every state in the game is visited infinitely often with probability 1 and hence the game is ergodic. 
	
\end{itemize}

\smallskip\noindent \textbf{Proof of Theorem \ref{theo:ds}.}
We have already shown that the game is ergodic. The rest is obtained from the modeling and the determinacy result.

\subsection{Formal Modeling of Proof of Stake Pool Attack} \label{app:model_stake}
\begin{itemize}
	\item {\em Game States.}
	We consider two pools, $A$ and $B$ and, as in the block withholding game, assume that any miner outside the two pools mines independently and set $\epsilon = \frac{1}{2n+1}$. We set $S\subseteq \{1, 2, \ldots, n\} \times \{1, 2, \ldots, n\} \times \{0, 0.01, 0.02, \ldots, 1\}$, where each state in $S$ is of the form of a 3-tuple like $s = (i, j, p)$ and corresponds to a situation in the game where pool $A$ has a total mining power of $i \epsilon$, pool $B$ has $j \epsilon$ and whenever a mined block is announced, the number of independent stake signatures that it receives is drawn from the Poisson distribution with parameter $(1 - i \epsilon - j \epsilon) p$. The intuition is that $p$ is a measure of connectivity of the network and each miner sees the block and signs it with this probability. We are using Poisson distribution as a rough continuous approximation of the binomial distribution. In real life, the distribution can be obtained by trial and error on the network.
	\item {\em Actions at each state.}
	Each pool has two choices at each state: to sign a block mined by the other pool, or to refrain from signing. We show these with $a_1^s, a_1^r$ for pool $A$ and $a_2^s, a_2^r$ for $B$.
	\item {\em Rewards.}
	We consider pool $A$'s revenue as game rewards. Several cases should be considered:
	\begin{enumerate}
		\item If $A$ is chosen to mine the next block and the mined block gets signed by a majority of stakes, either including $B$ or not, then $A$ gets a mining reward of $10$ units.;
		\item If $B$ is chosen to mine the next block and $A$ opts to sign it then $A$ gets a signing reward of $i \epsilon$, i.e. the total signing reward for each block is $1$ unit;
		\item Similarly if an independent miner, or $A$ itself for that matter, gets to mine the next block, $A$'s revenue will be $i \epsilon$ units \footnote{We assume that $A$ always signs blocks found by itself and independent miners.}.
	\end{enumerate}

	More concretely, we have $R_A = R_1 + R_2 + R_3$, where $R_A$ is the revenue of pool $A$ and $R_i$ corresponds to revenues from each of the parts above. Let $CDF$ denote the cumulative distribution function corresponding to the distribution mentioned above, then we have:
	\begin{itemize}
		\item $R_1 =10  i \epsilon \times \left\{\begin{matrix}
		1 & ~~&i \epsilon \ge \frac{1}{2}\\
		1 & ~~& \text{B chooses } a_2^s \text{ and } i\epsilon + j \epsilon \ge \frac{1}{2}\\
		1 - CDF(\frac{1}{2}-i\epsilon) &  ~~&\text{B chooses } a_2^r \text{ and } i \epsilon < \frac{1}{2}\\ 
		1 - CDF(\frac{1}{2}-i\epsilon-j\epsilon) & ~~& \text{B chooses } a_2^s \text{ and } i\epsilon + j\epsilon < \frac{1}{2} 
		\end{matrix}\right., $
		
		The first case corresponds to the situation where $A$ has enough stakes to sign his own block with a majority. The second case is when $A$ and $B$ form a majority together and $B$ has chosen to sign $A$'s block. In the third case, $A$ is not holding a majority and $B$ is not signing the block, so in order for $A$ to get the block mining fees, a fraction of other miners holding at least $\frac{1}{2} - i \epsilon$ must sign the block. The fourth case captures the state where both $A$ and $B$ sign the block but they do not form a majority.
		\item 
		$ R_2 = j \epsilon \times \left\{\begin{matrix}
		0 & ~~& \text{A chooses }a_1^r\\ 
		i \epsilon & ~~& \text{A chooses }a_1^s
		\end{matrix}\right.,$
		\item $R_3 = i \epsilon (1 -  j \epsilon).$
	\end{itemize}

	Pool $B$'s revenue, $R_B$, can be defined similarly and will be used in the next part.
	\item {\em Game transitions.}
	The attractiveness of a pool is defined as its revenue divided by its stake, i.e. $attr_A = \frac{R_A}{i \epsilon}$ and $attr_B = \frac{R_B}{j \epsilon}$. We do not consider the attractiveness of independent mining in this game. 
	A pool gains or loses mining stake based on its attractiveness. If it is the most attractive of the two, it gains $\epsilon$ stake with probability $\frac{2}{3}$, retains its current stake with probability $\frac{1}{6}$ and loses $\epsilon$ stake with probability $\frac{1}{6}$. Otherwise, it loses $\epsilon$ stake with probability $\frac{2}{3}$ and retains and gains with probability $\frac{1}{6}$ each. This is very similar to the case in the block withholding game.
	
	The value of $p$ remains the same or switches to one of the neighboring values with equal probability. This captures small changes in the network.
	
	\item {\em Ergodicity.} The argument for ergodicity is similar to the case of block withholding game. 
\end{itemize}

\smallskip\noindent \textbf{Proof of Theorem \ref{theo:ps}.}
As above, ergodicity of the game is established in the exact same manner as in the block withholding game. The rest follows straightforwardly from the modeling and the determinacy result.

%% file: app_exper.tex
\subsection{Details of Experimental Results} \label{app:experiment_details}
\textbf{Number of States.} The number of states in each of the experiments is determined as follows:
\begin{compactenum}
	\item \smallskip\noindent{\em Block withholding pool attack game.} In this game the number of states depends on the discretization factor of the mining power.
	For example, for a discretization factor $n$, 
	we say that pool A has $m$ units if its total mining power is $m/n$ fraction of the entire computation power of the network.
	Since we need to keep track of the mining power of A and B we have $O(n^2)$ states.
	\item \smallskip\noindent{\em Zero-confirmation double-spending game.}
	Here the number of states are exactly the different abstracted network states.
	For example if the minimal (maximal) odds for successful double-spending are 30\% (70\%) and we consider a discretization factor of $1/n\%$, then we will have $40n$ states.
	
	\item \smallskip\noindent{\em Proof of stake pool attack game.} Here the number of states is dependent upon both the discretization factor of mining stakes of the pools and the number of different abstracted network states. For example, if we consider $s$ network states and discretize mining stakes similar to Part 1 above, then the game will have $O(s n^2)$ states to keep track of the stakes of both pools and the connectivity of the network.
	
\end{compactenum}

\textbf{Experiment Machine and Parameters.} We obtained the results using an AMD Dual-core Opteron 885 (2.6 GHz) processor over Debian 3.2 OS with 32 GB of RAM and $\epsilon = 0.01$. The input is an ergodic game as described in Section \ref{sec:formal_short} and we are using Poisson distribution as the distribution mentioned in the formal modeling of the proof of stake pool attack game.